\documentclass[aps,pre,superscriptaddress,twocolumn,10pt,nofootinbib]{revtex4-1}
\usepackage{graphicx}
\usepackage{amsmath}
\usepackage{amsfonts}
\usepackage{amssymb}
\usepackage{color}

\def\wid{0.9}

\DeclareMathOperator\erf{erf}
\DeclareMathOperator{\upd}{d\!}

\begin{document}

\title{Pressure and Flow of Exponentially Self-Correlated Active Particles}

\author{Cato Sandford}
\affiliation{Center for Soft Matter Research and Department of Physics, New York University, 726 Broadway, New York, NY 10003, USA}
\author{Alexander Y. Grosberg}
\affiliation{Center for Soft Matter Research and Department of Physics, New York University, 726 Broadway, New York, NY 10003, USA}
\author{Jean-Fran\c{c}ois Joanny}
\affiliation{Physico-Chimie Curie UMR 168, Institut Curie, PSL Research University, 26 rue d'Ulm, 75248 Paris Cedex 05, France}
\affiliation{ESPCI-ParisTech, 10 rue Vauquelin 75005 Paris, France}

\date{\today}

\begin{abstract}
Microscopic swimming particles, which dissipate energy to execute persistent directed motion, are a classic example of a non-equilibrium system. We investigate the non-interacting Ornstein--Uhlenbeck Particle (OUP), which is propelled through a viscous medium by a force which is correlated over a finite time. We obtain an exact expression for the steady state phase-space density of a single OUP confined by a quadratic potential, and use the result to explore more complex geometries, both through analytical approximations and numerical simulations. In a ``Casimir''-style setup involving two narrowly-spaced walls, we describe a particle-trapping phenomenon, which leads to a repulsive effective interaction between the walls; while in a two-dimensional annulus geometry, we observe net stresses which resemble the Laplace pressure.
\end{abstract}

\maketitle


\section{Introduction}

Recent investigation of ``swimming'' particles has provided many new insights into non-equilibrium phenomena. These swimmers exhibit a persistent Brownian motion, which violates detailed balance and the fluctuation-dissipation theorem, and results in a range of behaviours not observed in passive systems \cite{E+G13,C+T15,F+M12,ESS13,LWG07}.

An ``Ornstein--Uhlenbeck Particle'' (OUP) swimmer is driven by a combination of a memory-less friction, and an exponentially correlated propulsion force with finite correlation time $\tau$. This model has already received significant attention, as it offers both a basic theoretical system for exploring non-equilibrium phenomena, and an accurate description of certain swimmer experiments \cite{Mag++14}. The OUP is furthermore closely related to two popular stochastic swimmer models (the active Brownian particle and the run-and-tumble particle), and complements them with different noise statistics.

Despite its relative simplicity, the OUP model is not generally solvable, and so a number of approximate methods have been developed to study their steady state densities -- for example the ``Unified Coloured Noise Approximation'' \cite{J+H87,Mag++15} or perturbative expansions close to equilibrium \cite{Fod++16, HDL89}.

In this paper we start with a simple exactly solvable model of an OUP confined in a one-dimensional harmonic potential, and discuss the crossover from an energy-equipartition dominated regime close to equilibrium, to a force-balance dominated regime far from equilibrium. We use the results to interpret simulation data on more subtle OUP interactions with external potentials, including flows generated by asymmetric potentials, attractive and repulsive Casimir forces and Laplace-like pressure on a curved surface.

Consider an OUP moving under an external force $\vec f(\vec x)$ arising from a potential $U(\vec x)$, $\vec f = - \nabla U$.  In one dimension (easily generalised to higher dimensions), the microscopic equation of motion for the OUP's coordinate $x(t)$ is the Langevin equation in which the propulsion force $\eta(t)$ plays the role of a coloured noise and has exponential correlations with a finite relaxation time $\tau$. To treat this problem, we imagine that fluctuations of $\eta(t)$ itself are governed by a hidden white noise variable $\xi(t)$, such that the system as a whole is described by coupled Langevin equations:
\begin{subequations}
	\begin{align}
		\zeta\dot x = \eta + f(x)
			\label{eqn:LE_dim_x} \\
		\tau\dot\eta = -\eta + \xi(t)
			\label{eqn:LE_dim_eta}
	\end{align}
	\label{eqn:LE_dim}
\end{subequations}
where $\left<\xi(t)\right>=0$ and $\left<\xi(t)\xi(t^\prime)\right>=2T\zeta\delta(t-t^\prime)$, with temperature $T$ (in energy units).  The amplitude of the correlation function is such that for a particle with no memory, $\tau=0$, the fluctuation-dissipation theorem is satisfied and equation~(\ref{eqn:LE_dim}) describes the dynamics of a passive Brownian particle, with equilibrium density determined by the Boltzmann distribution $\sim e^{-U(x)/T}$. The second equation ensures the exponential correlation of the propulsion force: $\left<\eta(t)\eta(t^\prime)\right> = \frac{T \zeta}{\tau} e^{-\left| t-t^\prime \right|/\tau}$.

The main novelty of our work is that of a \textit{method}: instead of viewing noise process $\eta$ as a nuisance to be integrated out as soon as possible, we retain this propulsion force as a phase-space variable. This enables calculation of phase-space currents and pressure formulae, on which all our results hinge.

The introduction of the hidden variable $\xi(t)$ allows us to recast the Langevin dynamics~(\ref{eqn:LE_dim}) in the form of a Fokker--Planck equation for the density $\rho(x,\eta)$:
\begin{align}
	\partial_t \rho = - \frac{1}{\zeta} \partial_x \left[ \left(\eta + f(x)\right) \rho \right] + \frac{1}{\tau} \partial_{\eta} \left[ \eta \rho \right] + \frac{\zeta T}{\tau^2} \partial^2_{\eta} \left[ \rho \right] \ .
		\label{eqn:FPE}
\end{align}
The first two terms on the right-hand side represent the advection in $x$ and $\eta$, and the last term is diffusion in $\eta$.


\section{Exact steady state}

Consider an OUP confined in a one-dimensional harmonic potential $U(x)=\frac{1}{2}kx^2$. The solution of the steady state Fokker--Planck equation~(\ref{eqn:FPE}) reads \cite{Sza14}
\begin{align}
	\rho(x,\eta) \propto \exp\left[-\frac{k}{2T}\left(\frac{k\tau}{\zeta}+1\right)\left[x^2 + \frac{k\tau}{\zeta}\left(\frac{\eta}{k}-x\right)^2\right]\right] \ ,
	\label{eqn:rho}
\end{align}
where $\frac{k\tau}{\zeta}$ is the dimensionless relaxation (or correlation) time.  The steady state currents in phase-space, according to Eq.~(\ref{eqn:FPE}), have components $j_x = \frac{1}{\zeta}(\eta-kx)\rho$ and $j_{\eta} = \frac{1}{\tau}\eta\rho + \frac{\zeta T}{\tau^2}\partial_\eta[\rho]$.  Current lines form closed loops on the $(x,\eta)$ plane, as shown in Fig.~\ref{fig:rho_E2} in appendix~\ref{app:DensityAndCurrents}.  While phase space loops in equilibrium systems may be observed for the pairs of phase-coordinates having opposite time-reversal signatures (such as position and velocity for an under-damped harmonic oscillator), our non-equilibrium system is different. The driving force $\eta(t)$, viewed as a phase-space variable, does not possess negative (velocity-like) time-reversal signature -- hence, this system violates detailed balance.

Integrating equation~(\ref{eqn:rho}) over all $\eta$ gives a Gaussian spatial density $n(x)$ with RMS displacement $\ell_{\mathrm OUP} = \sqrt\frac{T}{k}\left(\frac{k\tau}{\zeta}+1\right)^{-1/2}$, as has already been found by other means \cite{J+H87,Mag++15,Sza14}.  Thus, excursions of an OUP into the confining potential are smaller than those of its passive counterpart, $\ell_{\mathrm OUP} \leq \ell_{\mathrm passive} = \sqrt\frac{T}{k}$. This is the outcome of competition between two effects: more persistent particles explore the potential more efficiently, but at fixed temperature the increased persistence of $\eta$ is associated with a decreased amplitude
\footnote{\label{fnt:ABP-RTP} The ``temperature'' $T$ which appears in these equations was introduced in Eq.~(\ref{eqn:LE_dim_eta}) in order to construct the exponentially-correlated driving force; it may or may not have anything to do with the ambient temperature. Nevertheless, for our purposes it is natural to assume that $T$ is fixed, and thus the amplitude and correlation of $\eta(t)$ are simultaneously controlled by $\tau$}.

It is worth emphasising the physical origin of this penetration formula, which can be most easily apprehended by examining two limits. When $\frac{\tau k}{\zeta} \ll 1$ (close to equilibrium), the penetration is controlled by \textit{energy balance} $\frac{1}{2}kx^2 \simeq \frac{1}{2}T$.
In the opposite limit $\frac{\tau k}{\zeta} \gg 1$, it is controlled by \textit{force} $\eta \simeq kx$, such that the particle stalls when the
characteristic propulsion force $\eta =\sqrt{T\zeta/\tau}$ balances the potential force.

An active system's departure from equilibrium may also be identified with its rate of dissipation. For a quadratically confined OUP, it turns out that this dissipation is related to the OUP's average potential energy. To show this, we start with the equation of motion~(\ref{eqn:LE_dim_x}), multiply by a factor of $\dot{x}$ and average over time. The term which arises from the potential is a total time derivative, and vanishes in the steady state. Hence we are left with $\left< \zeta \dot{x}^2\right> = \left<\eta\dot{x}\right>$, which has a straightforward interpretation: the average power dissipated to friction equals the average power provided by the propulsion force.

The task now is to calculate what this power is in terms of the system parameters. Given the statistics of $\eta$, we may explicitly compute (see appendix~\ref{app:PowerBalance})
\begin{align}
	\left<\zeta\dot{x}^2\right> = \frac{1}{\tau}\frac{T}{1+\frac{k \tau}{\zeta}} \ .
		\label{eqn:DissipationRate}
\end{align}
The quantity $\frac{T/k}{1+\tau k/\zeta}$ is known to be equal to the mean-squared displacement of the OUP, so that $\frac{T}{1+\tau k/\zeta}$ may be thought of as an effective temperature (for a thorough discussion, see~\cite{Sza14} and also~\cite{Mag++14}). Equation~(\ref{eqn:DissipationRate}) therefore shows that an amount of energy equal to this effective temperature is dissipated on the correlation time-scale $\tau$. Put another way, equation~(\ref{eqn:DissipationRate}) becomes $\left<\zeta\dot{x}^2\right>=\frac{1}{\tau}\left<kx^2\right>$ -- the energy dissipated by the system in time $\tau$ is equal to twice the average potential energy.

Note that these calculations can be generalised to the case of a \textit{massive} particle propelled by and Ornstein--Uhlenbeck force (appendix~\ref{app:PowerBalance}); this yields further insights -- quantifying, for instance, the extent to which the Virial Theorem is violated.

\section{Pumping by an asymmetric potential}

We already noted the existence of currents in phase-space. Correlated dynamics may also produce currents in real space if they experience a potential landscape which breaks left-right symmetry -- something which has been observed in theoretical, experimental and biological systems \cite{Mag93,Kou++13,KML14, IBRR97,DLT97,Bar97}.
In principle, these currents offer a way of extracting work from systems of active swimmers.

\begin{figure}
	\centering
	\includegraphics[width=\wid\linewidth]{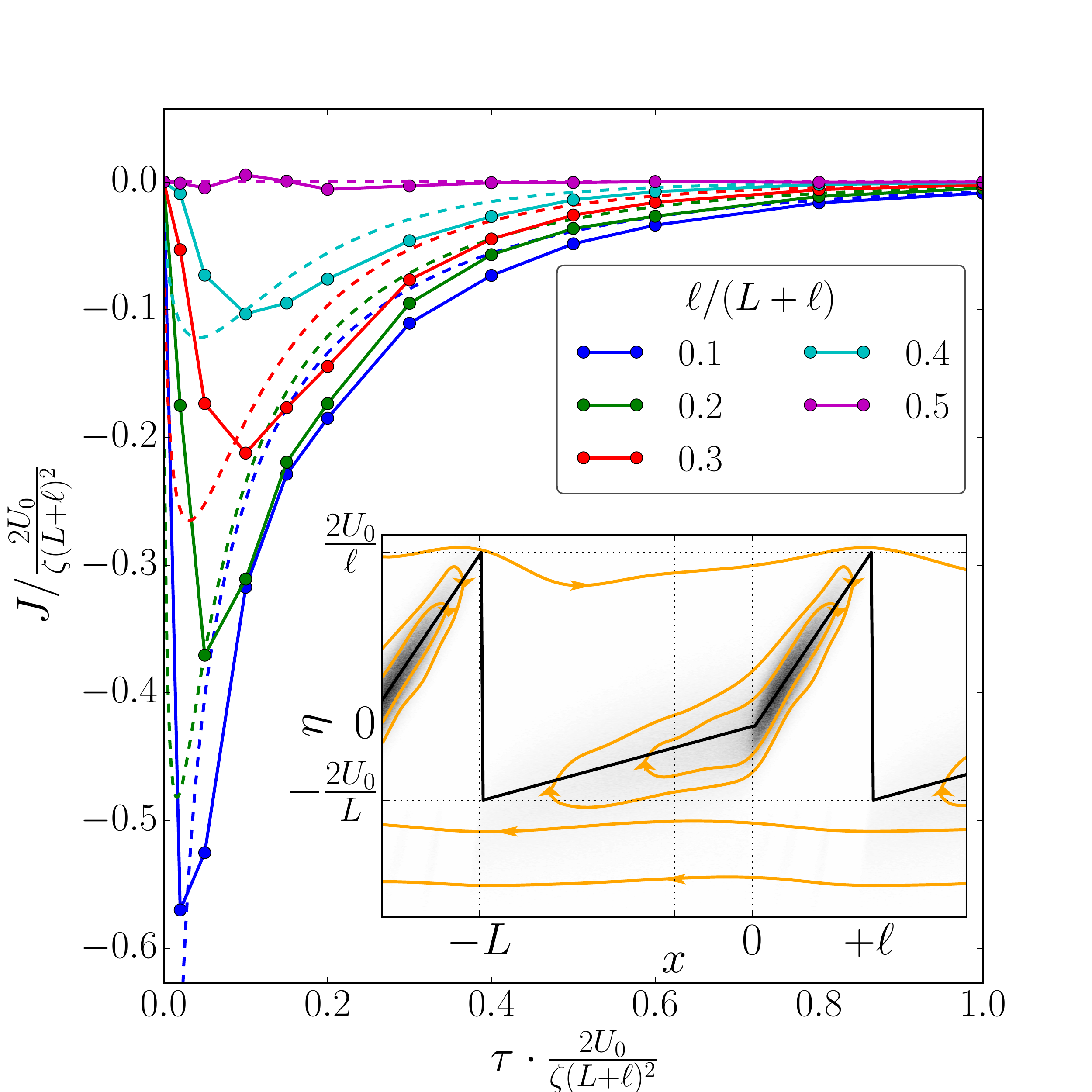}
	\caption{
		\textbf{Main figure:} Net current $J$ as a function of the correlation time (both measured in convenient units) for an OUP in a periodic, asymmetrical potential in 1D. Solid lines with markers show simulation results for several degrees of asymmetry, while dashed lines show the approximate prediction described in the text. For these data, the height of the potential $U_0/T=1$, meaning the approximation described in the text is not fully applicable: yet it still captures the general behaviour.
		\textbf{Inset:} Contours of phase-space density. The solid straight lines show $-f(x)$. Current-lines are sketched and adorned with arrows.}
	\label{fig:rectify}
\end{figure}

As a specific example, consider an OUP in a one-dimensional potential $U(x)$ which is piecewise quadratic, asymmetric and periodic. We define the potential landscape $U(x) = U_0\, x^2/L^2$ for $-L \leq x \leq 0$ and $U(x) = U_0\, x^2/\ell^2$ for $0 \leq x \leq \ell$, with period $L+\ell$.  Numerical results for this system are presented in Fig.~\ref{fig:rectify}; a subfigure illustrates the force landscape, which is more relevant than the potential landscape because, unlike the classical case of an \textit{energy} barrier, OUPs must overcome a \textit{force} barrier \cite{Mag93,DLT97}. Particles therefore move to the right (or left) on the $(x,\eta)$ plane only when $\eta > -f(x)$ (or $\eta < -f(x)$), since there is no diffusion along $x$, only drift.

These results can be understood quantitatively by considering the limit of small penetration into either side of the potential, such that the current along $x$ is small. In this case, we can use the density given by Eq.~(\ref{eqn:rho}). The total current in the $+x$ direction over the {force barrier} at $x= \ell$ is obtained by integrating the current $j_x$ over all $\eta$ larger than the force barrier $2U_0/\ell$. A similar calculation yields the current in the $-x$ direction, and the sum of these two contributions is the net current $J$.  This prediction compares reasonably well with simulations in Fig.~\ref{fig:rectify}.  As stated, this procedure is justified when the penetration depth is small compared to the sizes of the force barriers.  In this case, the overall current is also small.  We do not attempt in this work to analyse the applicability limits more accurately and to estimate the possible corrections.  We note nevertheless that, judging by our limited numerics, this approximation appears to hold qualitatively well beyond the low-current regime.

\section{Pressure}

Further consequences of the non-equilibrium character of OUPs can be found in their production of mechanical stresses. This idea was investigated already in \cite{Sol++15}, where it was found that the pressure exerted by an ideal gas of active Brownian particles depends on torques exerted on them by the confining potential.
{We here consider point-like particles, so torque is not an issue.}

Since every particle located at coordinate $x$ exerts a force $f(x)$ on the source of the potential $U(x)$, the total average force is obtained by integration of $n(x)f(x)$. We now show how this quantity is connected to the statistics of $\eta$.  We derive equations for the first and second moments of $\eta$ by multiplying Eq.~(\ref{eqn:FPE}) by the appropriate power of $\eta$ and integrating over all $\eta$ \cite{KML14}. This gives (for arbitrary spatial dimension and with summation over repeated indices):
\begin{subequations}
	\begin{align}
		f_{i}(\vec{x}) & = - \left< \eta_{i} \right>(\vec{x})  \label{eq:average_eta_equals_force_D} \\
		f_{j}(\vec{x}) n(\vec{x}) & = \partial_{x_{i}} \sigma_{ij}(\vec{x}) \ , \label{eq:equation_for_second_moment_of_eta_D} \\ \ \sigma_{ij}(\vec{x}) & = \frac{\tau}{\zeta} \left[ \left( \left< \eta_{i} \eta_{j} \right> - \left< \eta_{i}\right> \left< \eta_{j} \right> \right) n(\vec{x}) \right] \ , \nonumber
	\end{align}
\end{subequations}
where Eq.~(\ref{eq:average_eta_equals_force_D}) encapsulates the steady-state balance of propulsion an potential forces on a single OUP, and Eq.~(\ref{eq:equation_for_second_moment_of_eta_D}) encapsulates the net balance of stresses on the OUPs' medium.

If the potential $U(x)$ depends on one coordinate only, representing a ``wall'' of the container, then the pressure on this wall is obtained by line integration of $f(x)n(x)$ in the direction perpendicular to the wall -- i.e. along $x$:
\begin{align}
	P = -\int_\text{bottom of wall}^\text{top of wall} f(x)n(x) \upd x,
	\label{eqn:P}
\end{align}
where ``bottom of wall'' and ``top of wall'' enclose a region with nonzero $f(x)$.
In general, however, the right-hand side of equation~(\ref{eq:equation_for_second_moment_of_eta_D}) is not a potential vector field. This means the line integral~(\ref{eqn:P}) depends on the integration path, and the concept of pressure is ill-defined beyond simple planar or spherical geometries.
Yet it turns out that even in these situations there are interesting physical effects.

\begin{figure}[ht]
	\centering
	\includegraphics[width=\wid\linewidth]{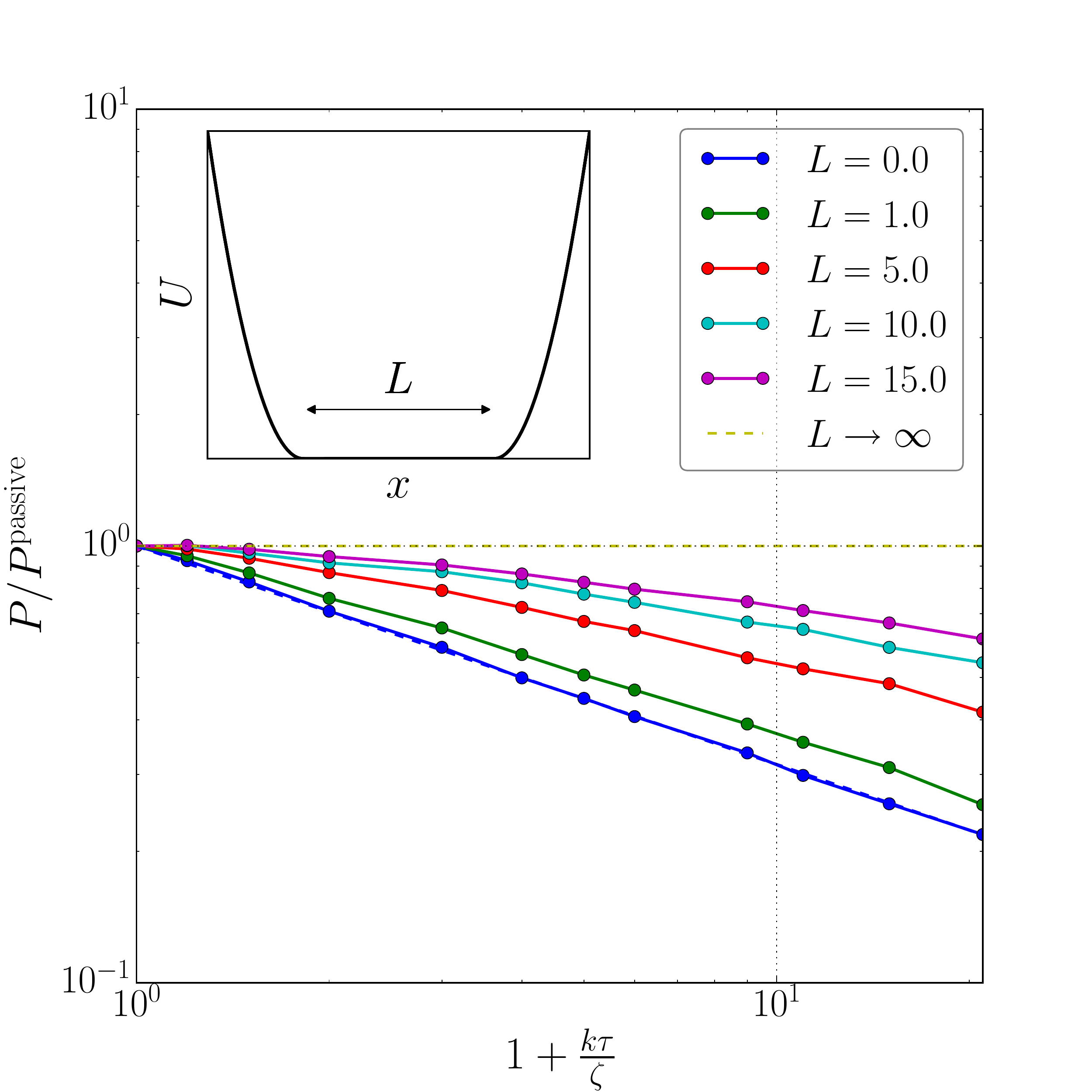}	
	\caption{Pressure as a function of the dimensionless correlation time $\frac{k\tau}{\zeta}$, for several bulk widths $L$. The prediction for $L=0$ is shown as a dashed line (obscured by data), and the prediction for $L\to\infty$ is a constant. The pressure exerted by an ideal gas of passive particles, $P^{\mathrm passive}$ is calculated by substituting the Boltzmann distribution into Eq.~(\ref{eqn:P}).}
	\label{fig:PA_CAR_DL}
\end{figure}

We begin by considering one-dimensional geometry, for which Eqs~(\ref{eq:equation_for_second_moment_of_eta_D}) and~(\ref{eqn:P}) imply the pressure on a wall $P = \frac{\tau}{\zeta} \left(\left[ \left< \delta \eta_x^2 \right> n(x) \right]_{\text{bottom}} - \left[ \left< \delta \eta_x^2 \right>  n(x) \right]_{\text{top}}\right)$, where $\left< \delta \eta_x^2 \right> \equiv \left<\eta_x^2 \right> - \left< \eta_x \right>^2$.
If the wall can be treated as infinitely high {{potential barrier}}, the second term contributing to the pressure vanishes.
Moreover, if there is a region between two confining walls where $f=0$ (as in the Fig.~\ref{fig:PA_CAR_DL} inset), the quantity $\left[\left< \delta \eta_x^2 \right> n \right]_{\text{bottom}}$ can be evaluated anywhere in this ``bulk''. Thus the pressure exerted on the walls depends solely {{on}} bulk quantities, and OUPs in 1D obey an equation of state.

We might imagine that when the width of the bulk, $L$, is much larger than the persistence length over which a free OUP loses its $\eta$ correlation, $\sqrt{\tau T/\zeta}$, particles leaving one wall forget its influence by the time they reach the other one. More quantitatively, one can show that the variance of the propulsion force far from any walls is $T\zeta/\tau$. Combining this with the expression for the pressure, we obtain the familiar ideal gas law $P=n_0 T$, where $n_0$ is the density evaluated deep in the bulk.


Thus, in the limit $L \to \infty$, memory-driven active particles are no different from passive particles. The opposite limit, $L\to 0$, can be taken from the exact solution above. Fig.~\ref{fig:PA_CAR_DL} shows numerical results for intermediate cases.

\section{Repulsive ``depletion'' forces in a Casimir potential}

\begin{figure}[ht]
	\centering
	\includegraphics[width=\wid\linewidth]{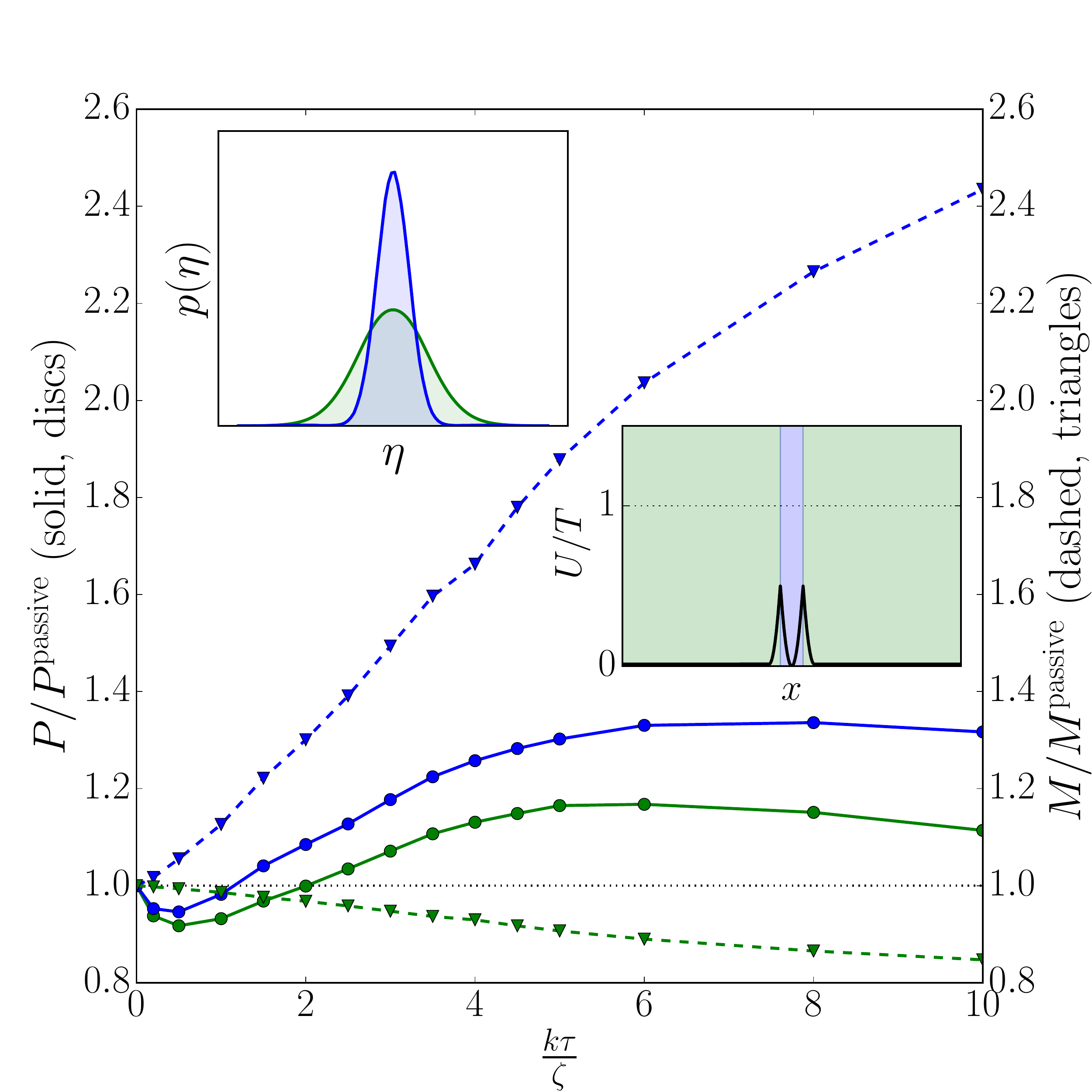}
	\caption{
	\textbf{Main figure:} {The main figure plots two sets of data. The pressure on the inner and outer portions of the Casimir potential (circles, solid lines, left ordinate axis), and the total probability of finding the OUP in each region (triangles, dashed lines, right ordinate axis), both as a function of the dimensionless correlation time $\frac{k\tau}{\zeta}$.}
	Here, ${P_\text{in}}>{P_\text{out}}$ and $M_\text{in}>M_\text{out}$; the height of the potentials is ${T/2}$ and their half-width is $\sqrt{{T}/{k}}$.
	\textbf{Upper inset:} A representative probability distribution of $\eta$ between the walls, which is narrower than the distribution in the {large bulk} ($p^\text{large}_\text{bulk}(\eta)\propto\exp\left[-\frac{\tau}{2T\zeta}\eta^2\right]$).
	\textbf{Lower inset:} Sketch of the piecewise-quadratic potential, whose the walls are penetrable for OUPs with sufficiently high $\eta$. For these data, the maxiumum height of the potential is $\frac{1}{2}T$, and the distance between the peaks is $2\sqrt{T/k}$.
	}
	\label{fig:PMa_Casimir}
\end{figure}

In this section, we consider a periodic ``Casimir''-style potential sketched in the lower inset of Fig.~\ref{fig:PMa_Casimir}. The potential consists of two narrowly-spaced walls, with a channel between them and a large (essentially infinite) bulk on either side. The walls themselves are permeable to OUPs which acquire sufficient propulsion to overcome the force barrier
\footnote{While the use of permeable walls is perhaps not typical for a Casimir experiment, and while they do change the physics of the situation slightly, the general thrust of the following discussion is not affected by them.}%
; and while we restrict ourselves here to 1D, similar results are obtained from analogous setups in higher dimensions.

We find numerically that the net pressure on the two interior walls does not in general vanish for OUPs: the solid lines in Fig.~\ref{fig:PMa_Casimir} {referring to the left ordinate axis} show they experience an effective repulsion. This is interesting because narrowly-separated walls typically \emph{attract}, due to the depletion of thermal or quantum fluctuations in the gap between them. The OUP case is different as a result of two competing effects.

To understand the first effect, consider an OUP in-between the two inner walls. If the gap is small, the particle does not have time to change its propulsion force $\eta$ before coming in contact with one of the walls. Particles with a large $\eta$ can cross the force barrier and escape, while particles with a small $\eta$ do not cross the force barrier and get trapped for at least a time $\tau$. As a consequence, the gap between the inner walls is populated mostly by lackadaisical particles and the probability distribution in the gap is strongly peaked around $\eta=0$. This is indeed observed -- see the upper inset of Fig.~\ref{fig:PMa_Casimir}, which compares the distribution of $\eta$ {between} the two interior walls with the distribution in a large bulk. This is somewhat analogous to the conventional Casimir effect, and it consequently \textit{lowers} the interior pressure $P_\text{in}$ relative to the exterior pressure $P_\text{out}$ (since low-$\eta$ particles don't penetrate far into the wall region).
Yet in Fig.~\ref{fig:PMa_Casimir}, we observe $P_\text{in}>P_\text{out}$ -- the walls \textit{repel} each other -- so this effect cannot be dominant.

Concomitant with the low magnitude of $\eta$ is a disproportionate \emph{accumulation} of particles in the region between the walls: once they reach this region, it is difficult for them to leave, because the narrowly-spaced walls constantly sap the particles' propulsion force.  {This is illustrated by the dashed lines referring to the right ordinate axis in Fig.~\ref{fig:PMa_Casimir}.} The narrow gap between interior walls therefore acts as a trap, concentrating the particle density and raising the pressure to an extent that outweighs the diminished penetration effect discussed in the previous paragraph. This effect has no analogy in the {regular} Casimir scenario.

To explore the physics further, we consider a slightly different periodic potential that is more amenable to explicit calculations. Similar to the original Casimir potential depicted in Fig.~\ref{fig:PMa_Casimir}, this new potential features two narrowly-spaced steep walls flanked by a broad region where the potential force is relatively small: therefore we may expect to see some of the same physics at play. The new potential is piecewise-quadratic, with one piece possessing smaller curvature than the other: $U(x)=U_0\left(\frac{x}{L}+1\right)^2$ for $-2L\leq x\leq 0$ and $U(x)=U_0\left(\frac{x}{\ell}-1\right)^2$ for $0\leq x\leq 2\ell$, with  $L\gg \ell$ ensuring that the second region is narrow compared to the first.  The period of $U(x)$ is then $2L+2\ell$ (see Fig.~\ref{fig:CasimirPotential} in appendix~\ref{app:CasimirModel} for an illustration).

At the steady state, the flux out of the narrow interior region is balanced by the flux into it, a fact which can be expressed as $M_{\mathrm in}k_{\mathrm in\to out}=M_{\mathrm out}k_{\mathrm out\to in}$ (where the $M$s are the total probability in the inner and outer regions, and the $k$s are rate constants). For this potential, $k_{\mathrm in\to out}$ and $k_{\mathrm out\to in}$ differ, because the \textit{height} of the force barriers and the force \textit{gradient} are both direction-dependent. This is similar to the particle-pumping potential in Fig.~\ref{fig:rectify}, and the difference between the rate constants can be investigated using the same machinery: choosing parameters such that the OUP penetration into any wall is relatively shallow, we use the density equation~(\ref{eqn:rho}) as an approximation for each potential well. Combining these densities with the zero-flux condition, we show in appendix~\ref{app:CasimirModel} that even for moderate values of $\tau$, OUPs are highly confined to the narrow region between the two walls, in agreement with Fig.~\ref{fig:PMa_Casimir}.

We stress that the potential used for this calculation is somewhat different from our original Casimir potential.
There, OUP accumulation between the walls was due to the reinforcement of correlations in
$\left< \eta^2 \right>$ by the proximity of the walls. In the case just considered the heights of the force barriers
are in addition direction-dependent. This scenario is therefore a little closer to the one considered in \cite{Sol++15},
where ABPs interacted with different potentials on either side of a hard piston.

The non-monotonicity of the OUP pressure exerted on the Casimir potential can be explained by a competition between varying penetration into the walls and enhanced accumulation between them. When $\frac{k\tau}{\zeta}$ increases from $0$, the pressure initially follows the average penetration and decreases below the thermal value. However, the force-controlled accumulation of particles with low $\eta^2$ begins to dominate around $\frac{k\tau}{\zeta}\gtrsim 1$. Finally, when $\frac{k\tau}{\zeta}$ is large enough that the penetration is smaller than the half-width of the interior wall, each region becomes increasingly isolated, and we are back to (multiple copies of) the situation in Fig.~\ref{fig:PA_CAR_DL}.


\section{Active Laplace pressure}

Interior walls are not the only way to break spatial symmetry and induce
pressure gradients. Swimmers interact with \emph{curved} walls in a nontrivial manner, as has
been observed in ABP simulations \cite{FBH14,S+L15,Nik++16} and experimental systems
\cite{Gal++07,PhysRevE.89.032720}. The simplest setup involving both positive and negative curvature,
but avoiding ambiguities in the definition of pressure, is an annular geometry.

\begin{figure}[ht]
	\centering
	\includegraphics[width=\wid\linewidth]{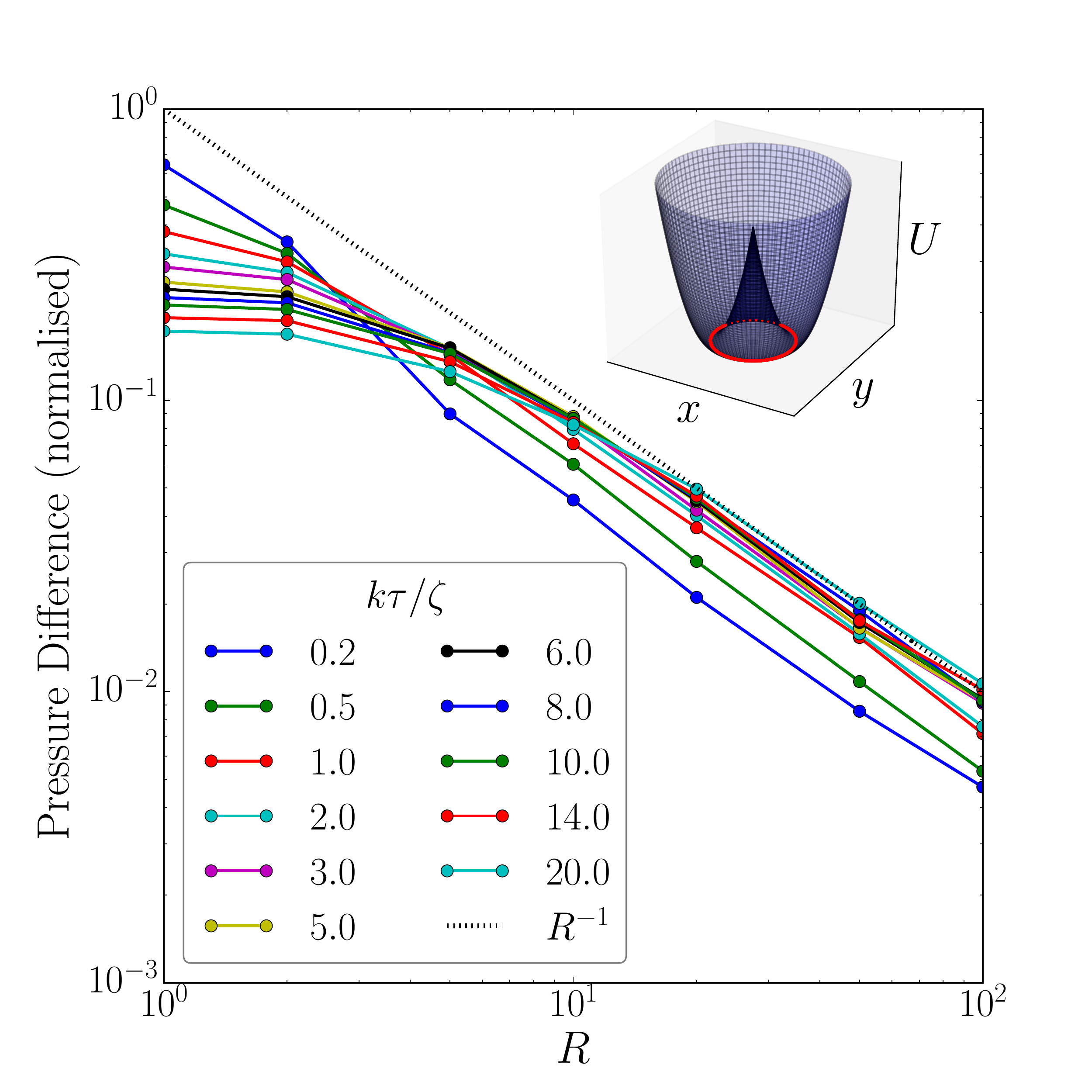}
	\caption{
	\textbf{Main figure:} The pressure difference $(P_\text{out}^\text{OUP}-P_\text{in}^\text{OUP})/(P_\text{out}^\text{flat}\sqrt\frac{k\tau}{\zeta})$ for an annular potential, as a function the wall position $R$ and for several values of the dimensionless correlation time $\frac{k\tau}{\zeta}$. (We divide the pressure difference by the pressure for a flat wall in order to fix normalisation as $R$ changes, and we also divide by the free-particle persistence length $\sqrt\frac{k\tau}{\zeta}$ for better comparison of curves.) The bulk is of zero width and located at $r=R$; and the line $1/R$ is indicated by dots.
	\textbf{Inset:} Schematic of the annular potential in 3D, with the foot of the wall indicated.}
	\label{fig:DP_polar}
\end{figure}

Even in this highly symmetrised setting, explicit results are forthcoming on neither the radial density profile nor the pressure on the inner and outer walls
\footnote{Previous approximate work on OUPs in a radially-symmetric geometry \cite{Mag++15} did accurately describe some phenomena, for instance that the probability distribution peak is offset towards regions of low curvature.}.
Thus, we examine numerically the statistics of an OUP confined in the potential $U(r)=\frac{1}{2}k(r-R)^2$, where $R$ is a
parameter which determines both the curvature of the annulus and the position of the (zero-width) bulk.  We observe that
OUPs tend to collect in the ``concave'' outer wall region (see movie in the supplementary material).
This is consistent with what has been found previously for simulations of ABPs confined by hard walls
\cite{FBH14,S+L15,Nik++16}, and is also intuitively reasonable: persistent particles in the inner convex region
may escape by changing their direction just a little (or not at all), while those in the concave outer region must make a
more drastic change to their direction to escape. The difference in density between the inner and outer regions leads
to a difference in pressure on the inner and outer walls, with $P_\text{outer}>P_\text{inner}$. Numerical results for
the pressure difference $\Delta P$ as a function of $R$ are plotted in Fig.~\ref{fig:DP_polar}. As expected, when
$R\to \infty$ and the curvature asymmetry between the walls vanishes, $\Delta P$ does too.  Moreover, when $R$ is
large enough to make the potential effectively infinite at $r=0$, we find $\Delta P\propto 1/R$. This is reminiscent
of a Laplace pressure, with effective surface tension depending on the dimensionless correlation time $\frac{k\tau}{\zeta}$.


\section{Concluding remarks}

In this paper we examined how non-equilibrium flows and pressure imbalances develop in systems of non-interacting particles driven by a stochastic correlated force, $\eta(t)$. The exact steady-state density $\rho(x,\eta)$ for a single OUP confined in a one-dimensional quadratic potential reveals two distinct regimes. \textit{Low} values of the dimensionless correlation time $\frac{k\tau}{\zeta}$ lead to an equilibrium-like regime of approximately passive particles, while \textit{high} values are associated with the balance between $\vec\eta$ and the potential force.

We show how potential barriers {and force barriers} influence the spatial distribution of OUP propulsion forces, and how this phenomenon can be exploited to produce net currents and unbalanced mechanical pressures.
In one dimensional simulations, two narrowly-separated walls (reminiscent of a Casimir setup) experience an effective repulsion. This arises because the potentials sap the particles' propulsion and act as traps. This phenomenon was further investigated with an analytic approximation, which gives similar results.
Curved boundaries also induce pressure imbalances. For propelled particles confined in an annular geometry, we find the difference in pressures on the outer and inner confining walls is proportional to the boundary curvature, as in Laplace's law.

\begin{acknowledgements}
	This work was supported primarily by the MRSEC Program of the National Science Foundation under Award Number DMR-1420073.  AYG acknowledges useful discussions with M.~Kardar.
\end{acknowledgements}


\appendix

\section{Exact solution for an Ornstein-Uhlenbeck Particle in a 1D quadratic potential}
\label{app:ExactSolution}

In this section, we shall assume length is meaured in units of $\sqrt{T/k}$, force in units of $\sqrt{Tk}$, and time in units of $\zeta/k$. We may then re-write the OUP model (Eqs~(\ref{eqn:LE_dim}) and~(\ref{eqn:FPE}) of the main text) in terms of the dimensionless correlation time $\alpha\equiv \tau k/\zeta$.

\subsection{Derivation of steady state density from Langevin equation}

Here we obtain equation~(\ref{eqn:rho}) of the main text directly from the (non-dimensional) stochastic equations. Combining equations~(\ref{eqn:LE_dim}) of the main text into a single vector equation for $\vec x\equiv\begin{pmatrix}x & \eta\end{pmatrix}^{\mathrm T}$:
\begin{align}
	\dot{\vec x} = A\vec x + \vec\xi(t) \ ,
	\label{eqn:LE_vec}
\end{align}
where $A=\begin{pmatrix}-1&1\\0&-1/\alpha\end{pmatrix}$ and $\left<\vec\xi(t)\vec\xi(t^\prime)\right>=\begin{pmatrix}0&0\\0&1/\alpha^2\end{pmatrix}\delta(t-t^\prime)$. Equation~(\ref{eqn:LE_vec}) can be ``solved'' as an integral over the stochastic force
\begin{align}
	\vec x(t) = \int_{-\infty}^{t} \exp[A(t-s)]\vec\xi(s) \upd s \ ,
\end{align}
and the covariance matrix $ C(t,t^\prime)\equiv \left< \vec x(t)\vec x(t^\prime)\right>$
\begin{align}
	C(t,t^\prime) = \int_{-\infty}^t \exp[A(t-s)]\left<\vec\xi(t)\vec\xi(t^\prime)\right>\exp[A^{\mathrm T}(t^\prime-s)] \upd s
\end{align}
which can be computed given the self-correlation of $\vec\xi$. Since equation~(\ref{eqn:LE_vec}) is a linear equation driven by a Gaussian process, its steady state density must be a bivariate Gaussian of the form $\rho(x,\eta)\propto\exp\left[-\vec x C^{-1}\vec x^{\mathrm T}\right]$. Performing the matrix exponentiation, multiplication and inversion, gives
\begin{align}
	\begin{split}
		\rho(x,\eta) = &\frac{\sqrt\alpha(\alpha+1)}{2\pi}\exp\left[-\frac{1}{2}(\alpha+1)^2x^2+\right.\\
			&\left.\quad-\frac{1}{2}\alpha(\alpha+1)\eta^2+\alpha(\alpha+1)x\eta\right] \ .
	\end{split}
\end{align}

\subsection{Density and currents in phase space}
\label{app:DensityAndCurrents}

\begin{figure}[ht]
	\centering
	\includegraphics[width=\wid\linewidth]{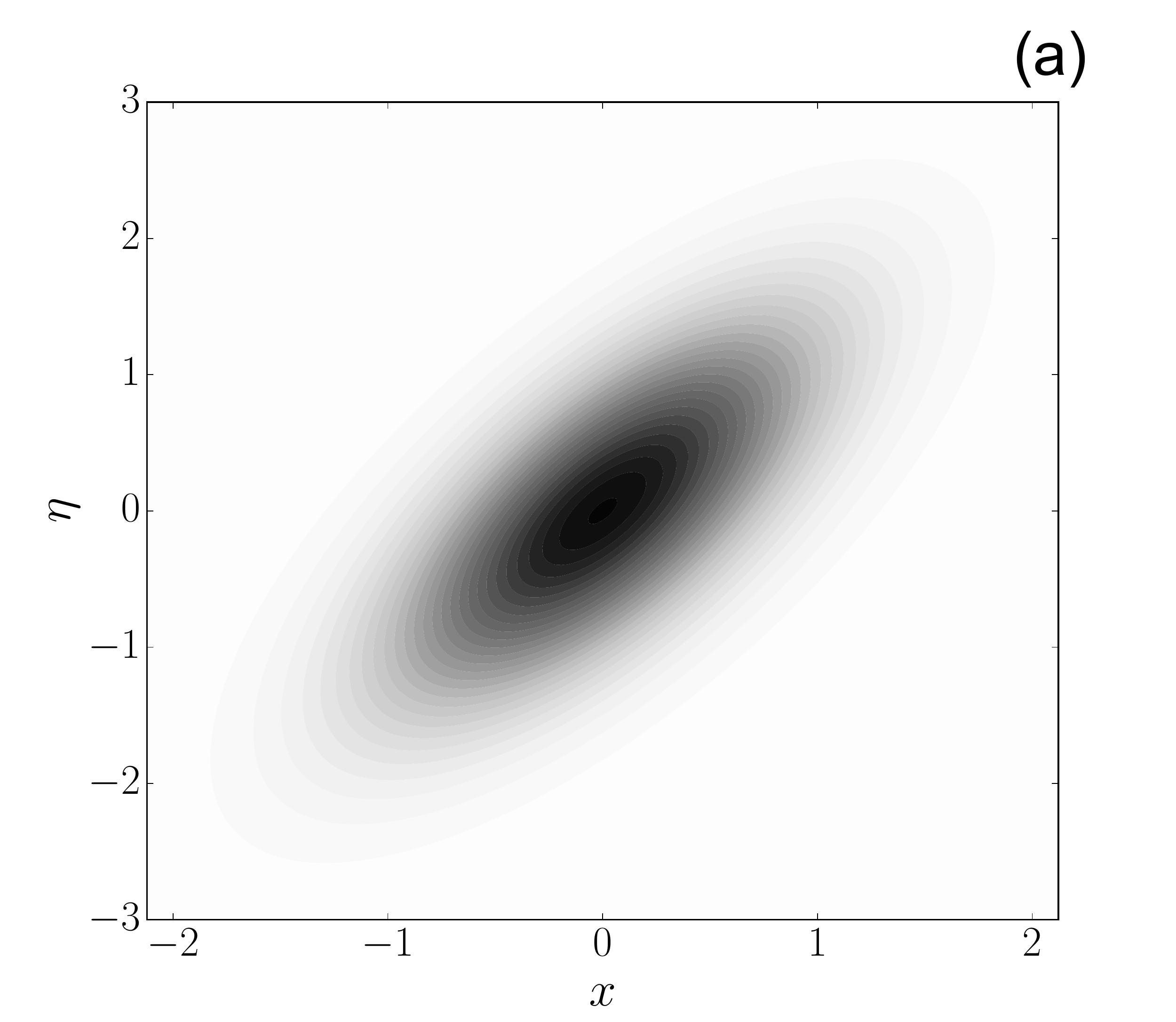}
	\includegraphics[width=\wid\linewidth]{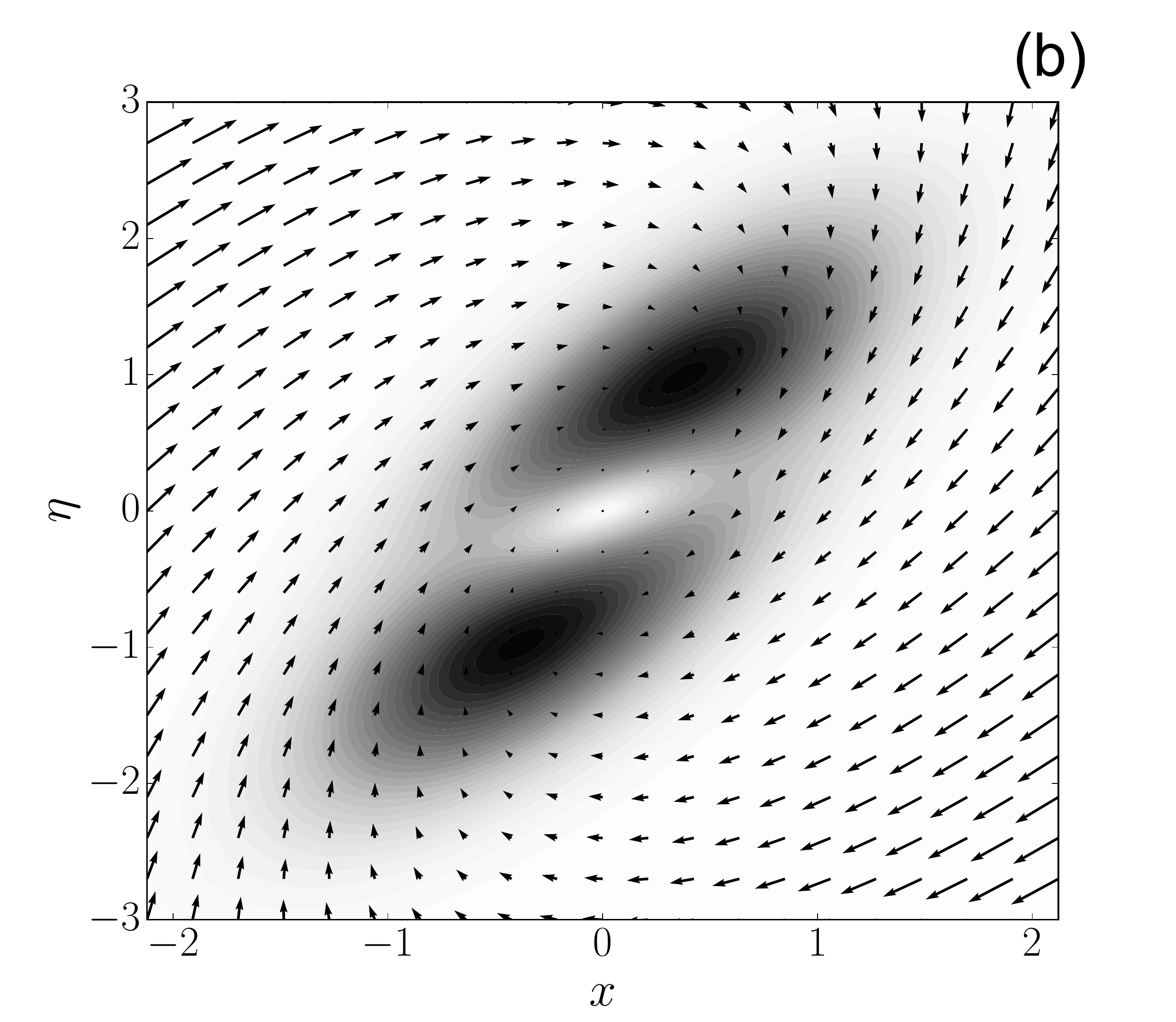}
	\caption{\textbf{Upper panel (a):} Density distribution in $(x,\eta)$ phase space.  Elliptical level lines illustrate the exact solution~(\ref{eqn:rho}).  \textbf{Lower panel (b):} Currents in $(x,\eta)$ phase space. Arrows represent \textit{velocity}, while the contours are magnitude of \textit{current}.}
	\label{fig:rho_E2}
\end{figure}

From equations~(\ref{eqn:FPE}) and~(\ref{eqn:rho}) in the main text, we find that steady-state currents exist in the full phase space, but cancel out when considering the $x$-coordinate alone (see Fig.~\ref{fig:rho_E2}).%

The spatial density $n(x)$ can be found from equation~(\ref{eqn:rho}) by integrating over $\eta$:
\begin{align}
	n(x) = \sqrt\frac{\alpha+1}{2\pi} \exp\left[-\frac{1}{2}(\alpha+1)x^2\right].
	\label{eqn:Q}
\end{align}
This exact solution, which agrees with approximations from the literature \cite{Mag++15,Fod++16}, has exponential form and hence can be mapped to a Boltzmann distribution by invoking an effective temperature
$T_{\mathrm eff}\equiv\frac{T}{\alpha+1}$ (in dimensionful units).

It is clear from the solution in equation~(\ref{eqn:rho}) that the level curves of the density in Fig.~\ref{fig:rho_E2} are concentric ellipses. Their eccentricity is
\begin{align}
	e = \sqrt{\frac{2 \sqrt{1 + 4 \alpha^2}}{1 + 2 \alpha + \sqrt{1 + 4 \alpha^2}}} \ .
\end{align}
This tends to unity in both $\alpha \to 0$ and $\alpha \to \infty$ limits, with a minimum of $e\approx 0.91$ at $\alpha=1/2$.

\subsection{Non-stationary mean-squared displacement}
\label{app:dispplacement}

From the overdamped Langevin equation, we can compute mean square displacement of an OUP.
Using units of $\sqrt{T/k}$ for $x$, $\zeta/k$ for time $t$, and with $\alpha = k \tau/\zeta$ being the dimensionless correlation time, we have
\begin{align}\begin{split}
	\left< \left[ x(t) - x(0) \right]^2 \right>  =  \frac{1 - e^{- \left( \alpha + 1 \right) t}}{ \alpha + 1 } +  \frac{1 - e^{ \left( \alpha - 1 \right) t}}{ \alpha - 1 }  e^{- 2 \alpha t} \ ,
	\label{eqn:MSD}
\end{split}\end{align}
with limits
\begin{align} \left< \left[ x(t) - x(0) \right]^2 \right> \simeq \left\{ \begin{array}{lcr} \alpha t^2 & \mathrm{for} & t \to 0 \\ \\ \frac{1}{\alpha + 1} & \mathrm{for} & t \to \infty \end{array} \right. \end{align}
The long time asymptotic corresponds to the confinement length which is implicit in the density distributions (\ref{eqn:rho}) or (\ref{eqn:Q}). The short time asymptotic, which is not diffusive but ballistic, reflects the fact that these particles are driven by the active propulsion force.

The relaxation time is controlled by the longer of the two time-scales in equation~(\ref{eqn:MSD}), namely $1/(\alpha+1)$ and $1/2\alpha$.

\section{Calculation of the Dissipation from a Quadratically Confined OUP}
\label{app:PowerBalance}

For a non-overdamped OUP of mass $m$, the equation of motion (ie, the balance of forces) reads
\begin{equation}
	m \ddot{x} + \zeta \dot{x} + kx = \eta \ , \label{eq:main_equation_only_colored}
\end{equation}
with $\left< \eta \right> =0 $ and $\left< \eta(t) \eta(t^{\prime}) \right> = 2 \zeta T \exp \left[ - \frac{\left| t - t ^{\prime} \right|}{\tau} \right] $ as before.  To obtain the balance of \textit{powers}, multiply both sides by $\dot{x}$ to arrive at
\begin{equation} \frac{\upd}{\upd t} \left[ \frac{kx^2}{2} + \frac{m\dot{x}^2}{2} \right] + \zeta \dot{x}^2 = \eta \dot{x} \ . \end{equation}
Averaging, we note that the first term vanishes in the steady state, so
\begin{equation} \left< \zeta \dot{x}^2 \right> = \left< \eta \dot{x} \right> \ , \label{eq:Power_Balance} \end{equation}
which simply means that the average power of dissipation by friction (the left hand side) is equal to the average power input provided by the propulsion force (the right hand side). We shall explicitly compute this power; but first we compute the mean squared-displacement for the massive OUP. Fourier transforming and performing a contour integral,
\begin{equation}
	\left< x^2(t) \right> \equiv \int_{-\infty}^{+\infty} \left( x^2 \right)_{\omega} \frac{\upd \omega}{2 \pi} = \frac{T}{k} \frac{1 + \frac{\tau \zeta}{m}}{1 + \frac{\tau \zeta}{m} + \frac{\tau^2 k}{m}} \ .
		\label{eqn:MSD_with_intertia}
\end{equation}
From this formula we recognise two familiar limits. For a system driven by white noise ($\tau \to 0$), $\left< x^2 \right> = T/k$ as required by equipartition. For the no-inertia case ($m \to 0$) considered in the main text we recover the previous finding $\left< x^2 \right> = \frac{T}{k} \frac{1}{1+ \frac{k \tau}{\zeta}}$.

The dissipation can be computed by considering either the right- or left-hand side of equation~(\ref{eq:Power_Balance}):
\begin{equation}
	\left< \zeta \dot{x}^2 \right> \equiv \int_{-\infty}^{+\infty} -\zeta \omega^2 \left(x^2 \right)_{\omega} \frac{\upd \omega}{2 \pi} = \frac{1}{\tau} \frac{T}{1+ \frac{k \tau}{\zeta} + \frac{m}{\tau \zeta}} \ .
		\label{eq:Power_General}
\end{equation}
Once again, two limits can be readily identified. When $\tau \to 0$, $\left< \zeta \dot{x}^2 \right> = \zeta T/m$, or $\left< m\dot{x}^2 \right> = T$, as expected from classical equipartition. For the no-inertia case, $m \to 0$, we arrive at equation~(\ref{eqn:DissipationRate}) from the main text.

Combining equation~(\ref{eq:Power_General}) with equation~(\ref{eqn:MSD_with_intertia}), we find a modified form of the Virial Theorem:
\begin{equation} \left( 1 + \frac{\tau \zeta}{m} \right) \left< \frac{m \dot{x}^2}{2} \right> = \left< \frac{k x^2}{2} \right>  \ .
	\label{eqn:ModifiedVirialTheorem}
\end{equation}
Deviations from the classical result are clearly parameterised by the non-equilibrium correlation time $\tau$.Taking the no-inertia limit of equation~(\ref{eqn:ModifiedVirialTheorem}) we find once more that the average potential energy is dissipated in time $\tau/2$.

\section{Calculations for OUP pumping in an asymmetrical potential}

We use formula~(\ref{eqn:rho}) of the main text as an approximation for the density.  In original units, we denote the un-normalised density in a quadratic potential with spring constant $\kappa$ as
\begin{align}
	\begin{split}
		& p_{\kappa}(x,\eta) \equiv \\ & \exp \left[ - \frac{\kappa}{2T} \left( \frac{\kappa \tau}{\zeta} + 1 \right) \left[ x^2 + \frac{\kappa \tau}{\zeta} \left( \frac{\eta}{\kappa} - x \right)^2 \right] \right]
	\end{split}\ . \label{eqn:standard}
\end{align}
Let the two different spring constants in the problem be
\begin{align}
	k = \frac{2 U_0}{\ell^2} \ \ \ \text{and} \ \ \ K = \frac{2 U_0}{L^2} \ , \label{eq:Definitions_of_k_and_K}
\end{align}
with $L\geq\ell$.
Then we approximate
\begin{align}
	\rho(x,\eta) \approx \left\{
		\begin{array}{lcr} A\, p_{K}(x,\eta) & \mathrm{for} & -L < x < 0 \\ & &  \\  a\, p_{k}(x,\eta) & \mathrm{for} & 0 < x < \ell \end{array} \right.
\end{align}
The ratio of the pre-factors $A$ and $a$ we fix by the (approximate) condition that the spatial distribution $n(x)$ is continous at the junction of the two potentials (at $x=0$):
\begin{align}
	A \int_{-\infty}^{+\infty} p_{K}(x=0,\eta) \upd \eta = a \int_{-\infty}^{+\infty} p_{k}(x=0,\eta) \upd \eta \ ,
\end{align}
yielding
\begin{align}
	\frac{A}{\sqrt{\frac{K \tau}{ \zeta}+1}} = \frac{a}{\sqrt{\frac{k \tau}{ \zeta}+1}} \ .
\end{align}
As a second condition, we assume (arbitarily) that the density is normalised in every period of the potential,
\begin{align}
	& \int_{-\infty}^{+\infty} \left[ A \int_{-L}^{0}  p_{K}(x,\eta) \upd x + a \int_{0}^{\ell} p_{k}(x,\eta) \upd x \right] \upd \eta = 1 \ .
\end{align}
Thus we obtain simple (but cumbersome) expressions for amplitudes $A$ and $a$.  We may then compute the current according to
\begin{align}\begin{split}
	J &=  a \int_{k\ell}^{\infty} p_{k}(x=\ell,\eta) \frac{\eta - k\ell}{\zeta} \upd\eta \ + \\
	&\quad + A \int^{-KL}_{-\infty}  p_{K}(x=L,\eta) \frac{\eta + KL}{\zeta} \upd\eta
\end{split} \label{eqn:current} \end{align}
where the first integral represents current to the right over the steep force barrier, and the second integral, which is negative, represents current to the left over the shallow force barrier.  In the end, dropping for clarity the normalization factor, one gets
\begin{align}
J \propto \frac{\exp \left[ - \frac{U_0}{T} \left(\frac{k \tau}{\zeta}+1 \right) \right]}{\sqrt{\frac{k \tau}{\zeta}+1}} - \frac{\exp \left[ - \frac{U_0}{T} \left(\frac{K \tau}{\zeta}+1 \right) \right]}{\sqrt{\frac{K \tau}{\zeta}+1}} \ .
\end{align}
Remembering definitions of spring constants $k$ and $K$, and letting $\ell = \lambda (L + \ell)$ and $L=(1-\lambda)(L+\ell)$, we finally arrive at
\begin{align}
J \propto \frac{\exp \left[ - \frac{U_0}{T} \left(\frac{ \alpha}{\lambda^2}+1 \right) \right]}{ \sqrt{\frac{\alpha}{\lambda^2}+1}} - \frac{\exp \left[ - \frac{U_0}{T} \left(\frac{\alpha}{(1-\lambda)^2}+1 \right) \right]}{ \sqrt{\frac{\alpha}{(1 -\lambda)^2}+1}} \ .
\end{align}
with dimensionless parameters in the problem being $U_0/T$ and $\alpha = 2U_0 \tau/(L+\ell)^2 \zeta$.  This current is plotted against $\alpha$ in Fig.~\ref{fig:rectify} of the main text, for various values of $\lambda$ (assuming $U_0/T = 1.0$ as an example).

\section{Approximation for the Casimir Potential}
\label{app:CasimirModel}

\begin{figure}[ht]
	\centering
	\includegraphics[width=\wid\linewidth]{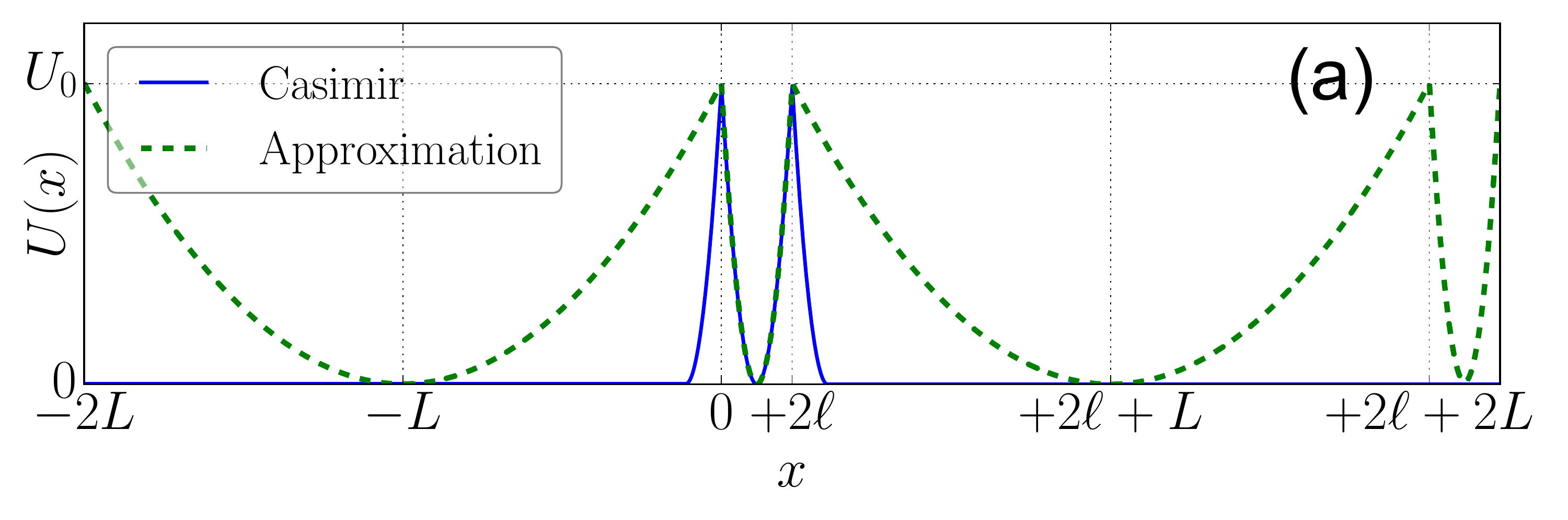}
	\includegraphics[width=\wid\linewidth]{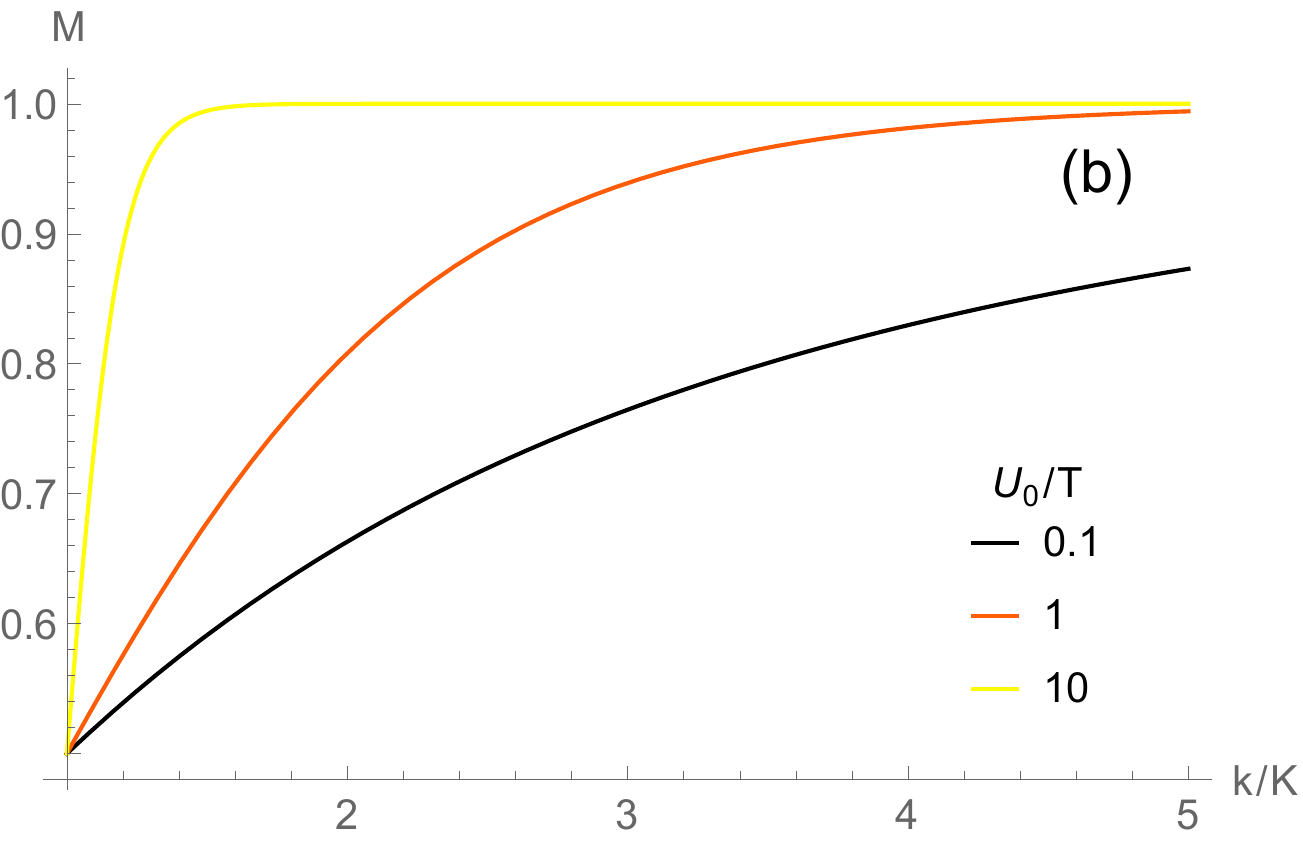}
	\caption{\textbf{Upper panel (a):} Schematic of the original Casimir potential from the main text (solid blue line), and the approximation to it considered here (green, dashed). \textbf{Lower panel (b):} The mass $M$ in the narrow region of the Casimir potential, as a function of the potential-stiffness ratio $k/K$, for several values of the potential height $U_0/T$.}
	\label{fig:CasimirPotential}
	\label{fig:CasimirMass}
\end{figure}

In the main text, we described how the exact result for the OUP density in a quadratic potential may be used to gain some insight into the observed accumulation between narrowly-spaced walls. We consider the potential
\begin{align}
	U(x) =
	\begin{cases}
		\frac{1}{2}K(x+L)^2 &\quad \text{for } -2L\leq x\leq 0\\
		\frac{1}{2}k(x-\ell)^2 &\quad \text{for } 0\leq x\leq 2\ell \ ,
	\end{cases}
	\label{eqn:Casimir_U}
\end{align}
with $K$ and $k$ defined as in equation~(\ref{eq:Definitions_of_k_and_K}), and $U_0$ the height of the energy barrier. The upper panel of Fig.~\ref{fig:CasimirPotential} compares this potential with the original Casimir potential considered in the main text.

Using the notation of equation~(\ref{eqn:standard}), the density in either well is approximated as
\begin{align}
	\rho(x,\eta) \simeq
	\begin{cases}
		& A \cdot p_{K}(x+L,\eta) \\
			&\qquad\qquad \text{for } -2L\leq x\leq 0\\
		&a \cdot  p_{k}(x-\ell,\eta) \\
			&\qquad\qquad \text{for } 0\leq x\leq 2\ell \ ,
	\end{cases}
	\label{eqn:Casimir_rho}
\end{align}
where $A$ and $a$ are factors to be determined. In the steady state, the net current over the force barrier at $x=0$ must be zero.  Similar to equation (\ref{eqn:current}), this gives one condition between amplitudes $A$ and $a$:
\begin{align}\begin{split}
	0 &= A \int_{KL}^{\infty} p_{K}(L,\eta) \frac{\eta - KL}{\zeta} \upd\eta + \\
	&\quad + a \int^{-k\ell}_{-\infty}  p_{k}(\ell,\eta) \frac{\eta + k\ell}{\zeta} \upd\eta
\end{split} \label{eqn:current_Casimir} \end{align}
The second condition which fixes amplitudes $A$ and $a$ is the normalization:
\begin{align}\begin{split}
	1 &= A \int_{-2L}^{0} \int_{-\infty}^{\infty} p_{K}(x+L,\eta) \upd\eta \upd x  + \\
		&\quad+ a \int_{0}^{2\ell}  \int_{-\infty}^{\infty}p_{k}(x-\ell,\eta) \upd\eta \upd x  \ .
	\label{eqn:norm_Casimir}
\end{split}\end{align}

\begin{figure}[ht]
	\centering
	\includegraphics[width=\wid\linewidth]{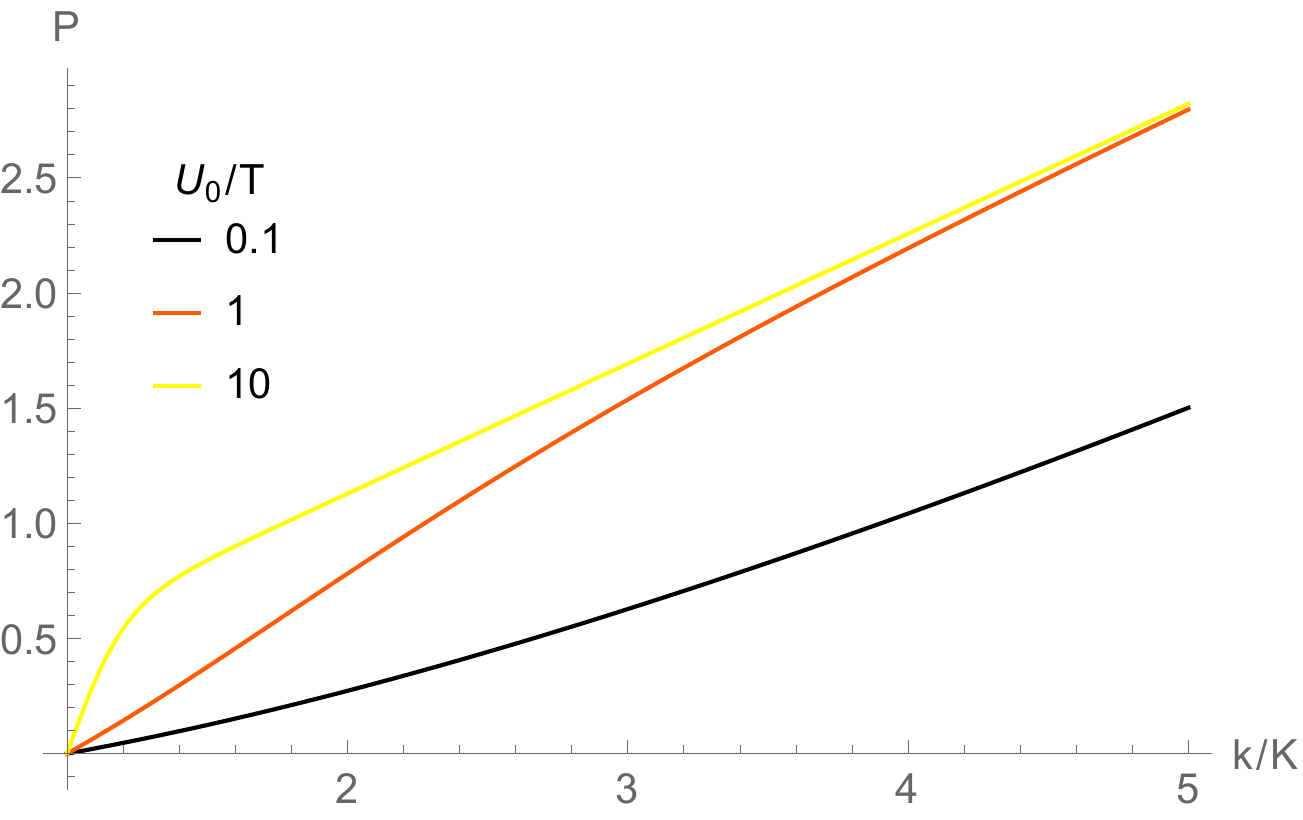}
	\caption{The net repulsive pressure $P$ on the interior walls of the Casimir potential, as a function of the potential-stiffness ratio $k/K$, for several values of the potential height $U_0/T$.}
	\label{fig:CasimirPressure}
\end{figure}

The total probability, $M$, to find the OUP in the narrow well can then be found as
\begin{align}
	M &= a \int_{-\infty}^{\infty} \int_{0}^{2\ell}  p_{k}(x-\ell,\eta) \upd x \upd\eta \\
	&= \frac{1}{ 1+ \left(\frac{ \frac{K\tau}{\zeta}+1}{\frac{k\tau}{\zeta}+1} \right)
		\frac{\exp\left[{\frac{U_0}{T}\frac{K\tau}{\zeta}}\right]}{\exp\left[{\frac{U_0}{T}\frac{k\tau}{\zeta}}\right]}
		\frac{\erf\left[\sqrt{\frac{U_0}{T}\left(\frac{K\tau}{\zeta}+1\right)}\right]}{\erf\left[\sqrt{\frac{U_0}{T}\left(\frac{k\tau}{\zeta}+1\right)}\right]} } \ .
		\label{eqn:CasimirMass}
\end{align}
This is plotted in Fig.~\ref{fig:CasimirMass}.

The total force on the wall is equal and opposite to the pressure. This is calculated as
\begin{align}
	\begin{split}
		-P &= A \int_{-L}^{0} K (x+L) \int_{-\infty}^{\infty} p_{K}(x+L,\eta) \upd\eta \upd x  + \\ &\quad+ a \int_{0}^{\ell} k (x-\ell)  \int_{-\infty}^{\infty}p_{k}(x-\ell,\eta) \upd\eta \upd x  \ . \label{eqn:pressure_Casimir}
	\end{split}
\end{align}
The expression is a little longer than equation~(\ref{eqn:CasimirMass}), so we merely plot it in Fig.~\ref{fig:CasimirPressure}.
Note that in this model, the net force exerted by OUPs on the walls always pushes them apart, as in the simulations of the Casimir potential in Fig.~\ref{fig:PMa_Casimir} of the main text.

\section{Position Trajectory for the Annular Geometry}

\begin{figure}[ht]
	\centering
	\includegraphics[width=\wid\linewidth]{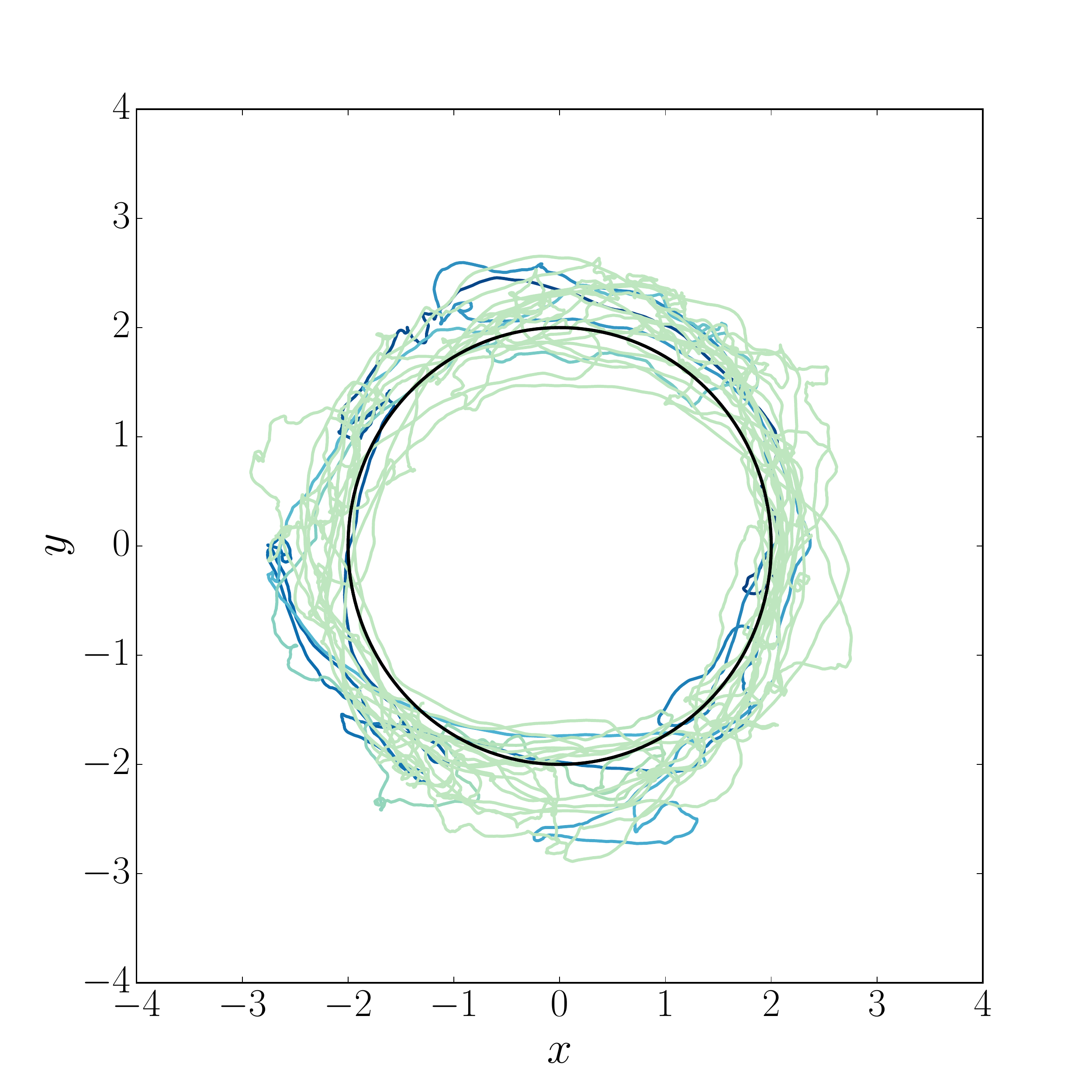}
	\caption{A sample trajectory trace in the annular geometry. Shaded according to time, with later times shaded darker. The solid black circle marks $r=R$.}
	\label{fig:traj_POL_DL_DR0}
\end{figure}

Fig.~\ref{fig:traj_POL_DL_DR0} shows a segment of an OUP trajectory trace in an annular potential with zero bulk. The trace is colour-coded according to time, with later times shaded darker. See also the movie in the supplementary material.

\section{Simulation Notes}
\label{app:Simulation}

Numerical simulation of equations~\ref{eqn:LE_dim} was implemented using an Euler--Maruyama scheme. The equations of motion were cast into dimensionless form using the prescription in appendix~\ref{app:ExactSolution}. (When there are multiple spring constants in the problem, we choose the largest, which gives the smallest unit of time.) We typically used the time-step $\Delta t=0.01$; although when the dimensionless correlation time $\alpha$ is small ($\alpha\lesssim 0.1$), it is prudent to employ a smaller increment in order to forestall issues with the convolution in constructing $\eta(t)$. Each data point shown here was generated from fifty runs of $100,000$ or so time-steps, which ensured that the steady state was reached and the initial condition had negligible influence on the final results. For simulations with relatively high force barriers, longer simulation times were occasionally needed to achieve steady state.

To avoid crowding in the plots of simulation results, we have omitted error bars. Deviation in the outcome of repeated runs was small, seldom more than the size of the plot markers.



\end{document}